\documentclass[aps,prb,twocolumn,superscriptaddress,showpacs]{revtex4}

\usepackage[dvips]{graphicx}
\usepackage{latexsym}

\begin{document}

\title{Specific heat and low-lying excitations
in the mixed state for a type II superconductor
}

\author{
N. Nakai
}
\email[]{nakai@yukawa.kyoto-u.ac.jp}
\affiliation{
Yukawa Institute for Theoretical Physics, 
Kyoto University, Kyoto 606-8502, Japan
}
\author{
P. Miranovi\'{c}
}
\affiliation{
Department of Physics, 
University of Montenegro, Podgorica 81000, Serbia and Montenegro
}
\author{
M. Ichioka
}
\affiliation{
Department of Physics, 
Okayama University, Okayama 700-8530, Japan
}
\author{
K. Machida
}
\affiliation{
Department of Physics, 
Okayama University, Okayama 700-8530, Japan
}

\date{\today}

\begin{abstract}
Low temperature behavior of the electronic specific heat $C(T)$ 
in the mixed state is by the self-consistent calculation 
of the Eilenberger theory. 
In addition to $\gamma T$-term ($\gamma$ is a Sommerfeld coefficient), 
$C(T)$ has significant contribution of $T^2$-term 
intrinsic in the vortex state. 
We identify the origin of the $T^2$-term as (i) V-shape density of states 
in the vortex state and (ii) Kramer-Pesch effect of vortex core shrinking
upon lowering $T$. 
These results both for full-gap and line node cases 
reveal that the vortex core is a richer electronic structure 
beyond the normal core picture. 
\end{abstract}

%
\maketitle

There has been considerable attention focused on the properties 
of a type II superconductor under an applied field,
ranging from conventional to exotic pairing superconductors,
such as cuprates.
Interest arises not only from a fundamental physics point of view,
but also from an application point of view,
such as the quest to achieve a higher critical current, etc.
Since practical applications of superconducting technology still
predominantly use ``conventional'' compounds, such as A-15 compounds;
Nb$_3$Sn or Nb$_3$Al as 
superconducting magnets,
it is of considerable importance to thoroughly understand the fundamental properties 
of the mixed state 
in an ordinary $s$-wave pairing state.
In spite of a long history\cite{fetter}, 
it is only recently that we are discovering the detailed electronic structures 
of the mixed state experimentally \cite{hess,sosolik,nishida,sonier} 
and theoretically\cite{ichioka,nakai}.
The latter has become possible because of combined efforts 
using both analytical and numerical computations based on a microscopic model.

The specific heat can be expanded as
\mbox{$C=\gamma T + c_2 T^2 +\cdots$} at a low temperature $T$ 
in the mixed state. 
The first term $\gamma T$ is the electronic contribution proportional to the 
zero-energy density of states (DOS) $N(E=0)$.  
The $\gamma$-term has been well studied for the purpose of determining  
the electronic state. For example, we can investigate the gap topology using 
the magnetic field $B$ dependence; 
$\gamma \propto B$ for $s$-wave full gap superconductor and 
$\gamma \propto \sqrt{B}$ for $d$-wave superconductor 
with line nodes in the superconducting gap~\cite{ichioka,nakai}.  
However, so far, the contribution of the $T^2$-term to $C$ 
has not been discussed~\cite{maki}.   
The purpose of this letter is to quantitatively estimate this $T^2$-term 
by microscopic calculation, 
and to clarify the existence and origin of the $T^2$-term. 
For this purpose, 
we also investigate the spectrum of 
the DOS $N(E)$ in the mixed state and 
$T$-dependence of the vortex structure.  
From this analysis we find that the $T^2$-term in $C$ comes from 
the electronic features intrinsic in the vortex state, 
i.e., V-shape DOS spectrum as seen later 
and the shrinkage of the core radius with decreasing $T$, namely,
the Kramer-Pesch (KP) effect~\cite{kramer,hayashi,ichiokaS}. 
When the vortex core radius keeps shrinking,   
$N(E=0)$ is expected to acquire an extra $T$-dependence. 
The KP effect is one of the consequences of the fact that the vortex can not be
pictured as a simple rigid core filled by normal electrons.  
The physics of vortices at low $T$ 
is much more interesting than the ``normal core model''.

Recently, several careful experiments on PbMo$_6$S$_8$\cite{junod},
Nb$_{77}$Zr$_{23}$\cite{junod}, 
$\beta$-pyrochlor CsOs$_2$O$_6$\cite{hiroi} and 
RbOs$_2$O$_6$\cite{hiroi} indicate a hint 
of a substantial deviation from the simple $\gamma T$ behaviors even at 
very low $T$. These materials are regarded as ``conventional'' superconductors.

Here we are going to numerically calculate $C(T)$, 
$N(E)$ and the vortex core size $\rho$ by solving the quasi-classical 
Eilenberger equation self-consistently 
with greater numerical precision than has hitherto been reported. Doing this is crucial for
examining the distinct,  
but subtle low $T$ behaviors of these quantities and allows us to find
the internal relationship between them. 
If lower numerical precisions are employed computational results yield only rounded features
that lead to the normal rigid core model.
We assume an isotropic $s$-wave superconductor in the clean limit.
To our knowledge, there has been no previous study on $C(T)$ and $N(E)$
to this level of accuracy\cite{fetter,maki,kusunose}.

%
%

The quasi-classical theory is valid for the case when \mbox{$k_{\rm F}\xi\gg 1$},
which is satisfied for almost all type II superconductors.
$k_{\rm F}$ is the Fermi-wave number 
and $\xi$ is the BCS coherence length which is our units of the length scale. We introduce the pair potential
\mbox{$\Delta({\bf r})$}, 
the vector potential 
\mbox{${\bf A}({\bf r})$}
and the quasi-classical Green's functions
\mbox{$g({\rm i}\omega_n, {\bf r}, \theta)$}, 
\mbox{$f({\rm i}\omega_n, {\bf r}, \theta,)$}
and \mbox{$f^{\dagger}({\rm i}\omega_n, {\bf r}, \theta)$}, 
where ${\bf r}$ is the center of mass coordinate of the Cooper pair.
The direction of the momentum,
\mbox{$\hat{\bf k}={\bf k}/\left|{\bf k}\right|$}= 
\mbox{$(\cos\theta, \sin\theta)$},
is represented by the polar angle $\theta$
relative to ${\bf x}$-direction.
The Eilenberger equation is given by
\begin{eqnarray}
\left\{\omega_n+\frac{\rm i}{2}{\bf v}_{\rm F}\cdot\left(\frac{\nabla}{\rm i}
+\frac{2\pi}{\phi_0}{\bf A}({\bf r})
\right)\right\}f
=\Delta({\bf r})g,
\nonumber 
\\ 
\left\{\omega_n-\frac{\rm i}{2}{\bf v}_{\rm F}\cdot\left(\frac{\nabla}{\rm i}
-\frac{2\pi}{\phi_0}{\bf A}({\bf r})
\right)\right\}f^{\dagger}
=\Delta^{\ast}({\bf r})g,
\label{eq1}
\end{eqnarray}
where
$g=[1-f^{\dagger}f]^{\frac{1}{2}}$, ${\rm Re}\, g>0$, 
\mbox{${\bf v}_{\rm F}=v_{\rm F}\hat{\bf k}$}
is the Fermi velocity, and 
$\phi_0$ is a flux quantum~\cite{ichioka,nakai}.
The applied field ${\bf H}$ is along the ${\bf z}$-direction.
In the symmetric gauge,
the vector potential is written as
\mbox{${\bf A}({\bf r})=
(1/2){\bf H}\times{\bf r}+{\bf a}({\bf r})$}, 
and the internal field ${\bf h}({\bf r})$ is given by 
\mbox{${\bf h}({\bf r})=\nabla\times{\bf a}({\bf r})$}. 

Equation (\ref{eq1}) can be self-consistently solved by numerical calculation, by considering 
the self-consistent conditions for
\mbox{$\Delta({\bf r})$} 
and
\mbox{${\bf a}({\bf r})$}; 
\begin{eqnarray}
\Delta({\bf r})
=N_0V_02\pi T\sum^{\omega_c}_{\omega_n>0}\int_0^{2\pi}\frac{d\theta '}{2\pi}
f({\rm i}\omega_n, {\bf r}, \theta '),
\label{scD} 
\\
{\bf j}({\bf r})=-\frac{\pi\phi_0}{\kappa^2\Delta_0\xi^3}
2\pi T\sum^{\omega_c}_{\omega_n>0}\int_0^{2\pi}\frac{d\theta}{2\pi}
\frac{\hat{\bf k}}{\rm i}g({\rm i}\omega_n, {\bf r}, \theta),
\label{scA} 
\end{eqnarray}
where 
${\bf j}({\bf r})=\nabla\times\nabla\times {\bf a}({\bf r})$ and 
$N_0$ is the density of states
at the Fermi level in the normal state.
The cut-off energy is set as \mbox{$\omega_{\rm c}=20T_{\rm c}$}.
\mbox{$\kappa=\sqrt{7\zeta (3)/72}
(\Delta_0/T_{\rm c})\kappa_{\rm GL}$}.
$\zeta$ is Riemann's zeta function.
The Ginzburg-Landau parameter is chosen as \mbox{$\kappa_{\rm GL}=9.0$}.
$\Delta_0$ is the uniform gap at $T=0$.

\begin{figure}[tb]
\includegraphics{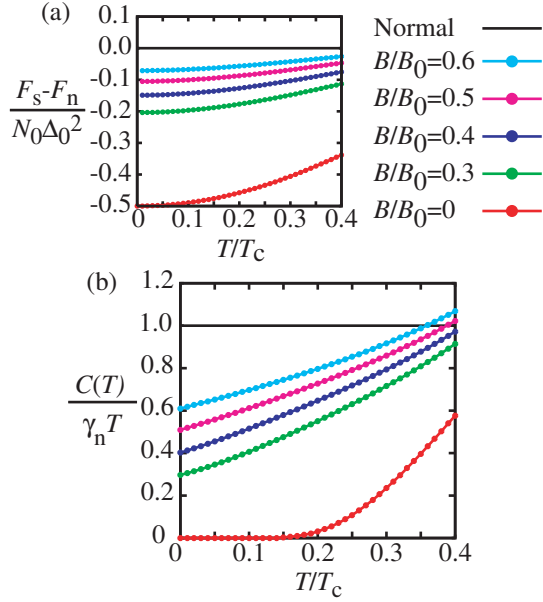}%
\caption{(Color online)
(a) Free energy difference $F_{\rm s}-F_{\rm n}$ and  
(b) specific heat $C(T)/T$ 
as a function of $T/T_{\rm c}$ for various fields 
$B/B_0=0, 0.3, 0.4, 0.5, 0.6$
from bottom to top.
\mbox{$B_{0}=\phi_0/\xi^2$}.
}
\label{fig:FC}
\end{figure}
Using the self-consistent solutions 
$g$, $f$, $f^\dagger$, $\Delta$ and ${\bf a}$,
the free energy difference is given by
\begin{eqnarray} && 
\frac{F_{\rm s}-F_{\rm n}}{N_0\Delta_0^2}=
\kappa^2
\frac{\langle |\nabla\times {\bf a}|^2 \rangle_{\bf r}}
{(\phi_0/\xi^2)^2}
\nonumber\\
 && \hspace{0.5cm} 
+\frac{2 \pi T}{\Delta_0^2}\sum_{\omega\geq 0}^{\omega_c} \left\langle
\int^{2\pi}_0 \frac{d\theta}{2\pi}
\frac{(g-1)(\Delta f^\dagger+\Delta^\ast f)}{2(g+1)}
\right\rangle_{\bf r} , \qquad 
\label{F} 
\end{eqnarray}
where
$F_{\rm s}(F_{\rm n})$ is the free energy for the superconducting (normal)
state, and $\langle\cdots\rangle_{\bf r}$ indicates the spatial average.
From $F_{\rm s}$, 
the specific heat coefficient $C/T$
is obtained by
\mbox{
$C/T=-\partial ^2 F_{\rm s}/\partial T ^2$
}.
The local density of states (LDOS) at an energy $E$
are given by
\begin{eqnarray}
N(E,{\bf r})=N_0 
\int_0^{2\pi}\frac{d\theta}{2\pi }
{\rm Re}\, g({\rm i}\omega_n\to E+{\rm i}\eta,{\bf r}, \theta),
\label{N} 
\end{eqnarray}
where $g$ is calculated by Eq. (\ref{eq1}) 
with \mbox{${\rm i}\omega_n\to E+{\rm i}\eta$}. 
We set \mbox{$\eta=0.01\Delta_0$}.
The total DOS $N(E)$ is the spatial average of the LDOS, i.e. 
$N(E)= \langle N(E,{\bf r}) \rangle_{\bf r}$. 
The self-consistent calculation is performed 
within the vortex lattice unit cell,
which is divided into \mbox{$81 \times 81$} mesh points.
We assume that the vortices form a triangular lattice.

The free energy difference $F_{\rm s}-F_{\rm n}$ 
and the specific heat $C(T)/T$ at low $T$ 
are shown in Fig. \ref{fig:FC} for various fields $B$,  
which are normalized by $B_{0}=\phi_0/\xi^2$.
We confirm that $C(T)/T$ in our numerical calculation at $B=0$ 
is identical with the BCS result. 
It is clearly seen that for $B \ne 0$, $C(T)/T$ exhibits 
a $T$-linear behavior at 
lower $T$ 
\mbox{($T/T_{\rm c}<0.2$)} where $C(T)$ for $B=0$
vanishes exponentially. 
Thus the specific heat in the mixed state 
can be expressed as 
\begin{eqnarray}
\frac{C(T)}{\gamma_{\rm n} T}=\gamma'+\alpha_c  \frac{T}{T_{\rm c}}+O\left(\left(\frac{T}{T_{\rm c}}\right)^2\right),
\label{C2} 
\end{eqnarray}
\noindent
at low $T$, where
$\gamma'=\gamma/\gamma_{\rm n}=N(E=0)/N_0$
and
$\gamma_{\rm n}=2\pi^2N_0/3$.
We also confirm that
\mbox{
$\gamma'(B)=\lim_{T\rightarrow 0} C/(\gamma_{\rm n}T)$}
shows a $B$-linear dependence expected for the $s$-wave pairing state\cite{nakai}.
It is remarkable to notice that $C/T$ is not constant at finite $T$, 
indicating a significant contribution from the $\alpha_c T$-term. 
When we estimate $\gamma'$ from $C/T$, 
the experimental data has to be carefully extrapolated toward $T \rightarrow 0$, 
in order to remove the $\alpha_c T$-term contribution.

\begin{figure}[tb]
\includegraphics[keepaspectratio]{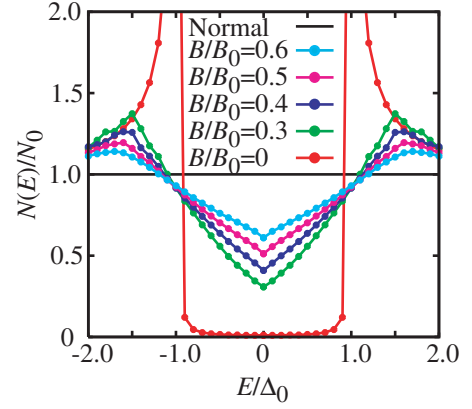}%
\caption{(Color online)
Energy dependence of the total density of states $N(E)$
is calculated at $T/T_{\rm c}=0.01$.
Plotted data are for $B/B_0=0, 0.3, 0.4, 0.5, 0.6$
from bottom to top at $E=0$,
showing  characteristic V-shape cusp structures
near the Fermi level $E=0$.
\label{fig:spc}
}
\end{figure}

\begin{figure}[tb]
\includegraphics[keepaspectratio]{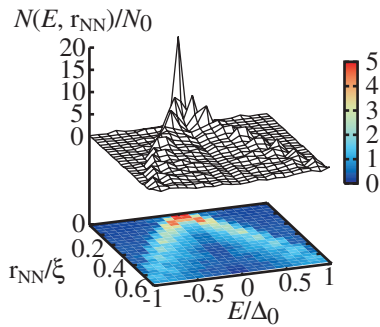}%
\caption{(Color online)
Spectral evolution
$N(E,{\bf r}_{\rm NN})$, 
where ${\bf r}_{\rm NN}$ is along the nearest neighbor vortex direction
of the hexagonal vortex lattice. 
The vortex center is situated at $|{\bf r}_{\rm NN}|=0$. 
$N(0,0)$ is truncated over 20.
\label{fig:ldos}
}
\end{figure}

The contribution of $\alpha_c T$-term in Eq. (\ref{C2}) 
also reveals important information on the electronic states. 
In order to analyze the origin of the $\alpha_c T$-term, 
we calculate the DOS spectrum and estimate the KP effect. 
The total DOS $N(E)$ is shown in Fig. \ref{fig:spc} 
at  our lowest $T (=0.01T_{\rm c})$.
We clearly see the V-shape DOS with a cusp structure for $B\ne0$. 
The low energy part of $N(E)$ can be fitted as
\begin{eqnarray}
\frac{N(E)}{N_0}=\frac{N(E=0,T)}{N_0}+\alpha_E \frac{|E|}{\Delta_0}.
\label{N2} 
\end{eqnarray}
As we show later more quantitatively, 
$N(E=0,T)$ and $\alpha_E$ correspond
to $T$-linear and $\alpha_c T^2$ contributions, respectively, 
to the specific heat $C(T)$.
From Fig. \ref{fig:spc}, we also see that 
$N(E=0) \sim \gamma$ 
increases with increasing $B$
and that $\alpha_E$ decreases as $B$ increases, so that the linear behavior of $|E|$
is unchanged. 

The $E$-dependence of $N(E)$ is related to the spectral evolution 
of the LDOS $N(E,{\bf r})$ around the vortex, 
which is shown in Fig. \ref{fig:ldos}. 
It is clear that the quasi-particle bound states 
have a characteristic dispersion relation $|E| \propto r$ 
as a function of the radius $r=|{\bf r}|$ from the vortex center, 
and form ridges extending outwards from the core.
The zero-energy peak at the vortex center gives rise to
\mbox{$N(E=0)$}, 
and the finite energy low-lying bound states near the core 
ultimately gives the $|E|$-dependent term in $N(E)$. 
A simple understanding of Eq. (\ref{N2}) at lower $B$ or lower $E$   
is as follows:
The total DOS can be evaluated by integrating LDOS spatially. 
Assuming circular symmetry around the core, 
$N(E)=N(E=0) +2\pi\int^\infty_0  
\delta(|E|-\beta r) r{\rm d} r$ per unit length along the field
direction.
The first term $N(E=0)$ mainly comes from LDOS $N(E=0,{\bf r})$ 
at the vortex core ${\bf r}=0$ (see the peak at $E=0$ in Fig. \ref{fig:ldos}). 
The second term comes from the quasi-particle
spectral weight whose spatial trajectory $|E|=\beta |{\bf r}|$, is indicated by 
the ridges in Fig. \ref{fig:ldos}.
This gives rise to V-shape total DOS: $N(E)=N(E=0)+2\pi |E|/{\beta^2}$.
These rich features are distinct from the simple normal core model,
which yields a featureless dispersion 
around the core. Some of the features of LDOS are directly observed 
in NbSe$_2$\cite{hess},
V$_3$Si\cite{sosolik} and YNi$_2$B$_2$C\cite{nishida}
via STM-STS experiments.

\begin{figure}[tb]
\includegraphics{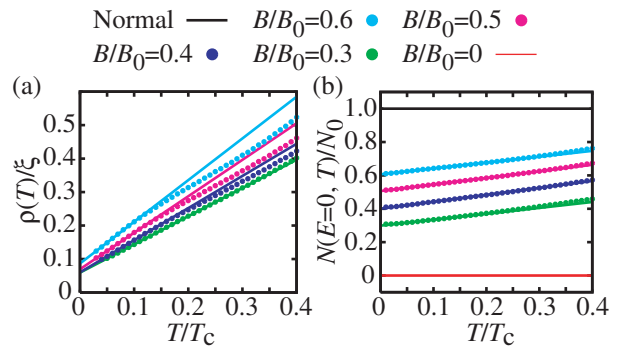}%
\caption{(Color online)
(a) Temperature dependence of the vortex-core radius $\rho$
and 
(b) Temperature dependence of the zero-energy density of states
$N(E=0,T)$
are plotted for $B/B_0=0, 0.3, 0.4, 0.5, 0.6$
from bottom to top. In Fig. (a) data for $B=0$ and normal state are not shown.
Lines are guides for the eye.
\label{fig:KP}
}
\end{figure}

We now estimate quantitatively the contribution from the KP effect to $C(T)/T$.
Here we define the core radius $\rho$ from the slope of $\Delta(r)$, i.e., 
\mbox{$\rho=\Delta_0 (\partial \Delta(r)/\partial r)^{-1}$} 
at the vortex core along the nearest neighbor vortex direction.
As shown in Fig. \ref{fig:KP}(a), 
$\rho$ decreases almost linearly on lowering $T$ 
for various values of $B$, which is a confirmation of the KP effect.
It is noted that the limit of $\rho$ when $T \rightarrow 0$ is finite
in the vortex lattice case,  
while $\rho \rightarrow 0$ at $T \rightarrow 0$ 
in the single vortex case~\cite{ichiokaS}. 
Since the zero-energy DOS due to the bound state around the vortex core 
is related to the core radius $\rho$, \mbox{$N(E=0, T)$} decreases 
as $T$ decreases, as shown in Fig. \ref{fig:KP}(b). 
This is approximately fitted by  
\begin{eqnarray}
\frac{N(E=0, T)}{N_0}=\gamma' + \alpha_{\rm K} \frac{T}{T_{\rm c}}.
\label{N02}
\end{eqnarray}
\noindent

\begin{figure}[tb]
\includegraphics[keepaspectratio]{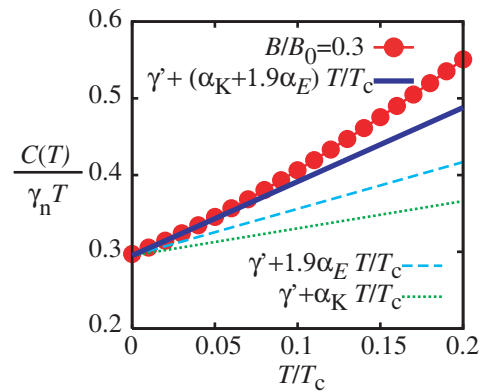}%
\caption{(Color online)
Specific heat $C(T)/T$ for $B/B_0=0.3$ (solid circles) with the estimated 
contributions  $1.9\alpha_E T/T_{\rm c} $,  $\alpha_{\rm K} T/T_{\rm c}$, 
and $(1.9\alpha_E + \alpha_{\rm K})T/T_{\rm c}$.
\label{fig:CT}
}
\end{figure}

Combining Eqs. (\ref{N2}) and ({\ref{N02}), we can express the DOS as
\mbox{$N(E, T)/N_0=\gamma'+\alpha_{\rm K} T/T_{\rm c}+\alpha_E |E|/\Delta_0$}. 
Through the relation 
\mbox{$C(T)/T=(2/T)\int_0^\infty dE E N(E, T) \partial f(E, T)/\partial T$}
using the Fermi distribution function $f(E, T)$ and
\mbox{$\Delta_0/T_{\rm c}=1.76$}, 
we obtain 
\mbox{$C(T)/(\gamma_{\rm n}T)=\gamma'+(\alpha_{\rm K} + 1.9\alpha_E) T/T_{\rm c}$}.
It is now clear that the $\alpha_c T $ term in $C(T)/T$ [Eq. (\ref{C2})] 
is affected by the KP effect term $\alpha_{\rm K} $ and $E$-linear
coefficient $\alpha_E $ equally.
To examine this relation, we analyze the data at $B=0.3$ as an example.
In Fig. \ref{fig:CT} we display $C(T)/T$ shown in Fig. \ref{fig:FC}(b), 
together with \mbox{$\gamma' +1.9 \alpha_E T/T_{\rm c}$} and 
\mbox{$\gamma'+(\alpha_{\rm K}+1.9\alpha_E) T/T_{\rm c}$}.
It can be seen that these two contributions  $\alpha_{\rm K}$ and $\alpha_E $ 
account for almost all $C(T)/T$ behavior at low $T$. 
That is, the KP term  $\alpha_{\rm K} $ and 
$E$-linear term $\alpha_E$ are both equally important for understanding
the total specific heat quantitatively.

We also study the $d$-wave pairing case with a line node, where 
$\Delta({\bf r}) \rightarrow \Delta({\bf r})\sqrt{2}\cos 2\theta$ 
in Eq. (\ref{eq1})~\cite{ichioka,nakai}, 
and obtain the same results as for the $s$-wave case:
V-shape DOS, KP effect and $\alpha_c T $-term in $C/T$\cite{anisotropy}. 
Note that $C(T)/T$ behaviors in Fig. \ref{fig:FC}(b) are qualitatively 
identical to the data of ${\rm Sr_2RuO_4}$ with a line node gap~\cite{deguchi}.

Our results of $\alpha_c T$-term in $C/T$ and V-shape DOS 
can be obtained after performing an exact calculation 
using the self-consistently obtained pair potential $\Delta({\bf r})$. 
When $C(T)/T$ and $N(E)$ are calculated by solving the same Eilenberger equation
within the Pesch approximation, 
it is found that there is neither a linear $\alpha_E |E|$ term 
in the DOS nor a linear term $\alpha_c T$ 
in the specific heat 
(see Figs. 1(b) and 1(d) in Ref. 14).

We discuss briefly the relationship between the V-shape DOS and physical 
quantities. 
The $|E|$-linear functional dependence in the V-shape DOS $N(E)$ happens to 
be the same as the DOS $N(E)\propto |E|$ for the line node gap structure 
at a zero field. 
From the $|E|$-linear dependence, 
we can evaluate the power law $T$-dependence
for various physical quantities by simple power-counting, such as 
specific heat ($C(T) \propto T^2$), 
nuclear relaxation time (\mbox{$T_1(T)^{-1}\propto T^3$}),
thermal conductivity $(\kappa(T)\propto T^2$), and ultrasound attenuation.
These behaviors are used to identify the line-node of 
the superconducting gap. 
In the mixed states, 
these power law components appear 
due to the V-shape DOS both for the full-gapped and line-node cases, 
in addition to the contributions by the zero-energy DOS $N(E=0)$. 
Therefore, in the experiment when a magnetic field is applied, 
we can not simply assign the origin of the power law behavior 
as a line node,   
since the origin may be the V-shape DOS due to the vortex states. 
The $T_1$-behavior $T_1(T)^{-1}\propto T^3$ under magnetic fields 
is a consequence of the spatial average. 
If we observe $T_1$ outside of the vortex core 
by using a site selective NMR technique\cite{takigawa}, 
we can unambiguously identify the signal due to the line node.

In summary, 
we have demonstrated that the electronic specific heat at low $T$
obeys the general law \mbox{$C(T)/T=\gamma'+\alpha_c T/T_{\rm c}$} 
in the mixed state of full-gapped superconductors. 
The contribution of second term $\alpha_c T/T_{\rm c}$, 
caused by the electronic features intrinsic in the vortex state, 
is significant in $C/T$ and observable. 
Correspondingly we have found the V-shaped DOS 
\mbox{$N(E)=N(E=0)+\alpha_E{|E|/\Delta_0}$} in the vortex states. 
The KP effect, shrinking vortex core radius with decreasing $T$, 
contributes to the $T$-dependence of the zero-energy DOS; 
$N(E=0, T)/N_0=\gamma' + \alpha_{\rm K} T/T_{\rm c}$.
We have shown that $\alpha_c \sim 1.9\alpha_E+\alpha_{\rm K}$, 
which shows the contributions from both the V-shape DOS and the KP effect 
are responsible for the $\alpha_c T/T_{\rm c}$ term in $C(T)/T$. 
We have also calculated for extreme anisotropic-gap cases, 
namely nodal gap cases, and found that the V-shape DOS 
and KP effect are not altered\cite{anisotropy}. 
We also pointed out some cautions when attempting to identify 
the gap topology experimentally, 
since the line node gap yields the same V-shape DOS even in the full gap
under a field.
 
We expect that careful measurement of $C(T)$ for clean type II superconductors
should confirm
our prediction. The V-shape DOS is directly observable by tunneling
experiments. These characteristics are found to be deeply rooted 
in the low-lying excitations around a vortex core, which 
goes beyond the rigid normal 
core picture, revealing a richer electronic structure in the vortex core.
We emphasize that the sample quality is important;
Impurity effects nullify the $T$-linearity $C(T)/T$.

\begin{acknowledgments}
We thank A. Junod and Z. Hiroi for helpful communications.
One of the authors (N. N.) is supported 
by a Grant-in-Aid for the 21st Century COE 
``Center for Diversity and University in Physics".
\end{acknowledgments}
%
\bibliography{basename of .bib file}

\end{document}